\documentclass[twocolumn,showpacs,prbprintnumbers,pre,fleqn,superscriptaddress]{revtex4}
\usepackage{epsfig}
\usepackage{graphicx}
\usepackage{amsfonts}
\usepackage[dvips]{color}
\usepackage{color}

\newcommand{\ct}{\cite}
\newcommand{\bi}{\bibitem}
\newcommand{\be}{\begin{equation}}
\newcommand{\ee}{\end{equation}}
\newcommand{\ba}{\begin{eqnarray}}
\newcommand{\ea}{\end{eqnarray}}
\newcommand{\al}{\alpha}

\newcommand{\ga}{\gamma}
\newcommand{\non}{\nonumber}
\newcommand{\bra}[1]{\langle #1|}
\newcommand{\ket}[1]{|#1\rangle}
\newcommand{\de}{\delta}
\newcommand{\noi}{\noindent}

\begin{document}
\title{ Study of Loschmidt Echo for a qubit  coupled to an  XY-spin chain  environment}
\author{Shraddha Sharma}
\email{shrdha@iitk.ac.in}
\affiliation{Department of Physics, Indian Institute of Technology Kanpur,
Kanpur 208 016, India}
\author{Victor Mukherjee}
\email{victor.mukherjee@cea.fr}
\affiliation{Institut de Physique Th\'{e}orique, CEA Saclay, F-91191 Gif-sur-Yvette Cedex, France}
\author{Amit Dutta}
\email{dutta@iitk.ac.in}
\affiliation{Department of Physics, Indian Institute of Technology Kanpur,
Kanpur 208 016, India}

\begin{abstract}
 
We study the temporal evolution of a central spin-1/2 (qubit) coupled to the environment which is chosen to
be a  spin-1/2 transverse XY spin chain. We explore the entire phase diagram of the spin-Hamiltonian and
investigate the behavior of Loschmidt echo(LE) close to critical and multicritical point(MCP). To achieve this, the qubit is coupled
to the spin chain through the anisotropy term as well as one of the interaction terms. Our study reveals that
the echo has a faster decay with the system size (in the short time limit ) close to a MCP and also  the scaling obeyed by the quasiperiod of the
collapse and revival of the 
LE is different in comparison to that close to a QCP.  We also show that even when approached along the gapless critical line, the scaling of
the LE is  determined by the MCP where the energy gap shows a faster decay with the system size. This claim is
verified by studying the short-time and  also the collapse and revival behavior of the LE at a quasicritical point on the ferromagnetic side
of the MCP. We also connect our observation to the decoherence of the central spin.
\end{abstract}

\maketitle

\section{Introduction}

Recent years have witnessed a tremendous progress in studies of quantum information theoretic
measures \ct{nielsen00,vedral07} close to a quantum critical point (QCP) \ct{sachdev99,chakrabarti96,continentino01}.
 Quantities like concurrence \ct{osterloh02,osborne02,amico08}, negativity \ct{peres96,vidal02}, 
quantum fidelity \ct{zanardi06,zhou08,gritsev10,gu10,rams10}, quantum discord \ct{sarandy09} etc., have been  found to capture 
the ground state singularities associated with a quantum phase transition (QPT); for recent reviews see \ct{dutta10,polkovnikov11}.

On the other hand, the studies of decoherence namely, the quantum-classical transition by a reduction
from a pure state to a mixed state  have also attracted the attention of physicists in recent
years \ct{haroche98,joos03,zurek91,zurek03}. In this connection, the concept of Loschmidt echo (LE) has been proposed
to describe the hypersensitivity of the time evolution of the  system to the perturbation
experienced by the environment to which it is coupled \ct{zurek94,peres95,jalabert01,karkuszewski02,cucchietti03}. The measure of the LE is
the modulus of the overlap between two states that evolve from the same initial state $|\psi_o \rangle$
under the influence of two Hamiltonians $H_0$ and $H_0 + \de$, where $\de$ is a small perturbation, given by

$$ L(t) = |\langle \psi_0| e^{i(H_0 + \de)t} e^{-iH_0t}|\psi_0\rangle|^2. $$

In some of the recent works, attempt has been made to connect these two fields by studying the behavior of the LE close to a QCP as a probe
to detect the quantum criticality. 
Quan $et~al$, studied the decay of LE using the central spin model where a central spin-$1/2$ (qubit) is coupled to the environment which  is
chosen to be a transverse Ising chain of $N$ spins in such a way that it is globally coupled to all the spins of the spin chain through the transverse
field term \ct{quan06}. The coupling to the qubit  leads to the perturbation term $\de$ defined above and consequently, the time evolution of the spin chain 
initially prepared in its
ground state, gets split in two branches both evolving with the transverse Ising Hamiltonian
but with different value of the transverse field.  This results in the decay in the LE. It has been observed that the LE shows a sharp decay in the vicinity of the quantum critical point of the environmental spin chain; at the same
time at the QCP, the LE shows collapse and revival as a function of time with the quasiperiod of revival of the LE being proportional to size
of the surrounding. This study has been generalized to the case where the environment is chosen to be a  transverse XY spin chain and the behavior of the
LE has been studied close to the Ising critical point driven by the transverse field \ct{yuan07,ou07}. 

Rossini $et~al$ \ct{rossini07}, studied a generalized central spin
model in which the qubit interacts with a single spin of the environmental transverse Ising spin chain and 
it has been shown that the decay of the LE at short time is given by the Gaussian form
$\exp(-\Gamma t^2)$ where the decay rate $\Gamma$ depends 
 on the symmetries of the phases around the critical point and the critical exponents.
 For instance, for such systems with local coupling, it has also been reported  that  $\Gamma$ has a  singularity
in its first derivative as a function of the transverse field at the QCP \ct{rossini07}. 
In a subsequent work \ct{zhang09}, the LE has been used as a probe to
detect QPTs experimentally; at the same time, using a perturbative study in the short-time
limit, the scaling relation  $\Gamma \sim (\lambda)^{-2 z \nu}$ valid close to a QCP (at $\lambda=0$) has
been proposed. Here, $\nu$ and $z$ are associated correlation length and dynamical
exponents, respectively \ct{sachdev99}. In contrast to these studies where the coupling
between the qubit and the environment is chosen to be weak, it has been shown that in
the limit of strong coupling the envelope of the echo becomes independent of the coupling strength which may arise due to quantum phase transition in 
the surrounding \ct{cucchietti07,cormick08}. Moreover the LE and the decoherence of the central spin has been studied when the environmental
transverse Ising spin chain is quenched across the QCP by varying the transverse field linearly in time \ct{damski11}.

\begin{figure}[ht]
\begin{center}
\includegraphics[width=8.5cm,height=6.0cm]{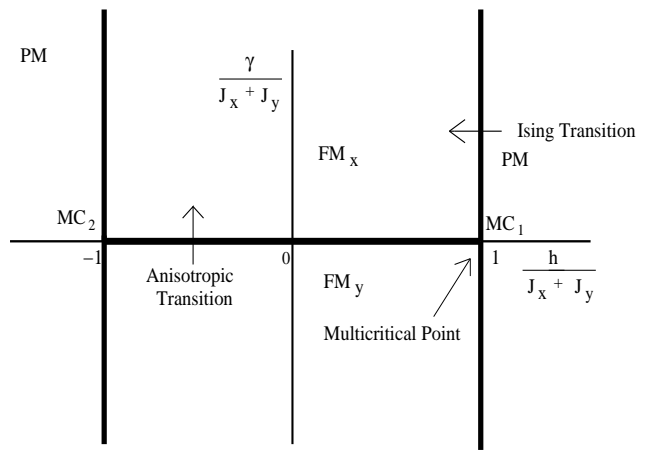}
\end{center}
\caption {The phase diagram of the anisotropic XY model in a
transverse field with Hamiltonian given by (\ref{hjx1}) in the $h/(J_x + J_y) - \gamma/(J_x+J_y)$ plane, where $\gamma=J_x -J_y$. The vertical bold lines denote Ising transitions from the ferromagnetic
phase  to the paramagnetic phase ($\rm{PM}$), whereas the horizontal bold line stands for the anisotropic phase transition between
 two ferromagnetic phases $FM_x$ and $FM_y$. The multicritical points at $J_x = J_y$ and $h = \pm1$ are denoted by $\rm{MC_1}$ and $\rm{MC_2}$, respectively.}
\label{Fig:XYphase}
\end{figure}

 The central spin model  we consider here, consists of a two level central spin $S$ coupled to an environment 
 $E$ which is chosen to be a  spin-$1/2$ $XY$ spin chain with anisotropic
interactions and subjected to a transverse field, described by the Hamiltonian
\be
H_{E} =- \sum_{i=1}^N[J_{x} \sigma^x_i \sigma^x_{i+1} + J_{y} \sigma^y_i \sigma^y_{i+1} + h\sigma^z_i],  \label{hjx1}
\ee
where $\sigma_{i}^\alpha (\alpha = x, y, z) $ are Pauli spin matrices,
 $h$ is the transverse field and $N$ is the total number of spins in $E$. 
The spin chain (\ref{hjx1}) is exactly solvable using Jordan-Wigner mapping from spins to fermions \ct{lieb61,barouch70,kogut79,bunder99}.
The phase diagram is shown in the Fig.~(\ref{Fig:XYphase}); the transition from the ferromagnetically ordered phase to the paramagnetic phase driven by the
transverse field $h$ is called the  Ising transition and the transition between two ferromagnetically ordered phase, with
magnetic ordering in the $x$ direction ($FM_x$) and the $y$ direction ($FM_y$), respectively, driven by the anisotropy parameter $\gamma=J_x -J_y$, is called
the anisotropic
transition. The anisotropic transition lines extend from $h=-(J_x + J_y)$ to $h=J_x+J_y$ along the  $\ga=0$ axis. Both Ising and anisotropic  critical lines meet
at   the multicritical points as shown in the figure. We shall exploit the exact solvability of the $XY$ spin chain to calculate the LE close to 
these critical points. We also
note at the outset that the spin chain is to be studied under periodic boundary condition
and the wave vector $k$ takes discrete values $k=2\pi m/N$ with $m=1, 2, ...N/2$
and the lattice spacing is set equal to unity.

Earlier studies focussed on the case when the coupling of the central spin   to the environment is through the transverse field $h$ \ct{quan06,yuan07,ou07} 
and explored the behavior of LE close to the Ising critical point. Motivation behind the
present work is to explore the short-time behavior, collapse and revival  of the LE
around the anisotropic critical point (ACP) and especially the multicritical point (MCP) $\rm{MC_1}$ of the phase
diagram. To achieve this we evaluate the LE by coupling the qubit to the anisotropy
term and also one of the interactions of  the spin chain. 
Finally, we  conjecture a generic scaling form  that should be valid close to
a QCP at least in the short-time limit. 

In the next section (Sec. II), we consider the case when qubit is coupled to
the anisotropy term and the behavior of the LE is explored;  in Sec. III, the qubit is coupled to  one of the interaction terms ($J_x$). Finally in the concluding section, we 
discuss our results and conjecture some generic scaling relations.

\section{Qubit coupled to the anisotropy term  of environment Hamiltonian}

In this section, it would be useful to  rewrite the Hamiltonian as
\ba
H_{E} = - \frac{1}{2}\sum_{i=1}^N[(1+\gamma) \sigma^x_i \sigma^x_{i+1} + (1- \gamma) \sigma^y_i \sigma^y_{i+1} 
+ 2h\sigma^z_i],
                                  \label{hE}
\ea
with the choice $J_{x}+J_{y} = 1$, and the anisotropy parameter $\ga=J_{x}-J_{y}$. 
Denoting the ground and excited states of the central spin  by $\ket{g}$ and $\ket{e}$, respectively,
the coupling of the system $S$ to the environment $E$ can be chosen as
\ba
H_{SE} &=& -  \frac{\delta}{2} \ket{e} \bra{e} \sum_{i=1}^N[ \sigma^x_i \sigma^x_{i+1} -  \sigma^y_i \sigma^y_{i+1} ], \label{hint}
\ea
where one assumes that  the excited state of the qubit couples to all the spins of the environmental spin chain.
The Hamiltonian of the composite system  ($S+ E$) is then given by
\ba
H_{e} &=& H_{E}+H_{SE} \label{hT}
\ea
 The form of the interaction Hamiltonian chosen in Eq.~(\ref{hint}), enables
one to analytically calculate the behavior of the LE close to the anisotropic transition line and also the MCP.
 
Let us assume the spin  $S$ to be initially in a pure state, $\ket{\phi(0)}_{S}=c_g\ket{g}+c_e\ket{e}$, (with coefficients satisfying $|c_g|^2+|c_e|^2=1$) and 
the environment $E$ be in the ground state denoted by  $\ket{\varphi (0, \ga)}_{E}$;  
the total wave function of the composite system at time $t=0$ can then be written in the direct product form
\ba
\ket{\Psi(0)}=\ket{\phi(0)}_S \otimes\ket{\varphi(0, \ga)}_E      \label{hwf1}
\ea
 One finds that the evolution of the $XY$ spin chain  splits into two branches (i) $\ket{
 \varphi (t, \ga) }=\exp(-iH (\ga) t)\ket{\varphi (0, \ga)}$ and
 (ii)$\ket{\varphi (t,\ga+\de) }=\exp(-iH (\gamma+ \delta) t)\ket{\varphi (0, \ga)} $; this implies that  $\ket{\varphi(t, \ga)  }$ evolves with the Hamiltonian
(\ref{hE}) with the anisotropy parameter $\ga$ and $\ket{\varphi(t,\ga+\de)}$ evolves  with the same Hamiltonian but the anisotropy parameter modified 
to $\gamma + \delta$.  The  total wave function at an instant $t$ is then given by
 \ba
\ket{\Psi(t)}&=&c_g\ket{g}\otimes\ket{\varphi(t, \ga)}+c_e\ket{e}\otimes\ket{\varphi (t, \ga+\de)}. \label{hwf2}
\ea
One therefore finds the decay of the LE given by \ct{quan06}
\ba
&L&(\gamma, t)= |\langle{\varphi(t, \ga)}\ket{\varphi(t,\ga+\de)}|^2 \non\\
&=& |\langle \varphi(0, \ga)|\exp(iH(\ga)t) \exp(-iH(\ga+\de)t) | \varphi (0, \ga) \rangle |^2\non\\
&=& |\langle \varphi(0, \ga)| \exp(-iH(\ga+\de)t) | \varphi (0, \ga) \rangle|^2 \non,\\
\label{hLE}
\ea
where we have used the fact that the  $|\varphi (0) \rangle$ is an eigenstate of the
Hamiltonian (\ref{hE}) with anisotropy parameter $\ga$.

The Hamiltonian (\ref{hE}) can be exactly solved by Jordan-Wigner (JW) transformation followed by Bogoliubov transformations \ct{lieb61,barouch70,kogut79,bunder99} and can be written
in the form $H (\gamma + \delta) = \sum_k \varepsilon_k(\gamma+\delta) (A_k^{\dagger} A_k -1/2)$ and $H (\gamma) = \sum_k \varepsilon_k(\gamma) (B_k^{\dagger} B_k -1/2)$
where $A_k$'s and $B_k$s are Bogoliubov fermionic operators and
\be
\varepsilon_k (\ga+\de) = \sqrt{(h+\cos k)^2 + \{(\gamma +\delta)  \sin k\}^2} \label{hL3};
\ee
clearly, $\varepsilon_k(\ga) =\varepsilon_k (\ga+\de)$ with $\de=0$. 
Here we have considered periodic boundary condition, the wave vector $k$ takes discrete values $k=2\pi m/N$ with $m=1, 2, ...N/2$ ( N is assumed to be even and also lattice 
spacing is set equal to one).

In fact, under the JW transformation the Hamiltonian (\ref{hE}) gets reduced to direct product  of decoupled $2 \times  2$ Hamiltonians for each momentum $k$ which in
the basis $|0\rangle$ (vacuum state) and $|k, -k \rangle$ (two JW fermion state)  can 
written as

 \begin{equation}
 H_k (\ga)= 
 \left(
 \begin{array}{cc}
  h + \cos k & i\ga \sin k  \\
 -i\ga \sin k& -(h + \cos k)   \\
 \end{array}
 \right).
 \label{rdm}
 \end{equation} 
 In the current problem in which the LE is calculated as a function of $\ga$, one makes
 resort to a basis transformation to $|\tilde 0\rangle$ and $|1\rangle$, such that
 $$ |\tilde 0\rangle= \frac 1 {\sqrt 2} (|0\rangle + i |k, -k \rangle) $$
 $$ |1\rangle= \frac 1 {\sqrt 2} (|0\rangle - i |k, -k \rangle), $$ 
 so that the reduced Hamiltonian (\ref{rdm}) gets modified to 
 \begin{equation}
 H_k (\ga)= 
 \left(
 \begin{array}{cc}
    \ga \sin k & h + \cos k  \\
h + \cos k & -\ga \sin k   \\
 \end{array}
 \right).
 \label{rdm1}
 \end{equation}  
  The ground state  of $H(\ga)$ and $H(\ga + \de)$ can be written in the form
 $$|\varphi(\ga,0)\rangle  =   \cos \frac {\theta_k(\ga)}{2} |\tilde 0\rangle -i\sin \frac {\theta_k(\ga)}{2} |1\rangle$$
  $$|\varphi(\ga+\de,0)\rangle  =   \cos \frac {\theta_k(\ga+\de)}{2} |\tilde0\rangle -i\sin \frac {\theta_k(\ga+\de)}{2} |1\rangle,$$ 
 where  $\tan \theta_k (\ga+\de)= {(h+ \cos k)}/{\{(\ga +\de)\sin k\}}$ and $\tan \theta_k (\ga) 
 = \tan \theta_k (\ga+\de)|_{\de=0}$. 
 
The Bogolioubov operators are related to the JW operators through the relation \ct{lieb61}
 \be
 A_{k} = \cos\frac{\theta_k (\ga+\de)}{2} a_{k} -i\sin \frac{\theta_k (\ga+\de)}{2} a_{-k} ^\dagger. \label{eq_btojw}
 \ee
 where $a_k$s are the Fourier transform of the JW operators as derived through the
 JW transformations of spins. 
 Using Eq. (\ref{eq_btojw}), one can further arrive at a relation connecting the Bogolioubov operators
 \be
 B_{k}= \cos(\alpha_{k}) A_{k} -i \sin(\alpha_{k}) A_{-k} ^\dagger
 \label{eq_AtoB}
 \ee
 where,
$\alpha_{k}=[\theta_k(\ga) - \theta_k(\ga+\de) ]/2 $
 
 Noting the fact that $A_k |\varphi(\ga+\de,0)\rangle=0$ and $B_k |\varphi(\ga,0)\rangle=0$  for all $k$, one can use the Eq.~(\ref{eq_AtoB}) to establish a connection between the ground states $ |\varphi(\ga+\de,0)\rangle$ and $ |\varphi(\ga,0)\rangle$ given by
 
 \be
\ket{\varphi (\ga,0)}= \prod_{k>0}[\cos(\alpha_{k})+i\sin(\alpha_{k})A_{k}^\dagger A_{-k}^\dagger]\ket{\varphi (0,\ga+\de)} . \label{hGS}
\ee
Substituting Eq. (\ref{hGS}) to the Eq.~(\ref{hLE}), we find the expression for the LE given by

\be
L(\gamma, t)=\prod _{k>0} L_{k} =\prod_{k>0}[1-\sin^2(2\alpha_{k})\sin^2(\varepsilon_ k(\ga+\de) t)] \label{hl1}
\ee

\begin{figure}[ht]
\begin{center}
\includegraphics[width=8.5cm,height=7.0cm]{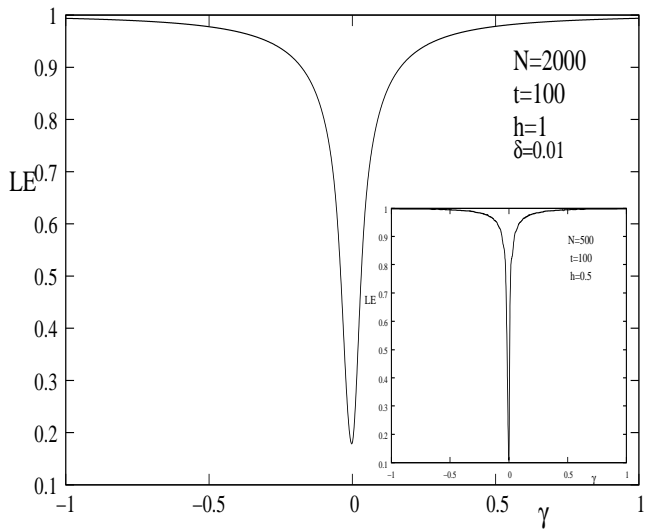}
\end{center}
\caption {The  LE as a function of $\gamma$ for   $h=1$; one observes a sharp dip
around the MCP ($\ga=-\de$). Inset shows a similar dip  around the aniosotropic critical
line (with $h=0.5$).}
\label{Fig:lemc}
\end{figure}

We shall use  Eq.~(\ref{hl1}) to calculate the
LE as a function of parameters $\ga$ and $h$, especially close the quantum critical points.
As  shown in Fig.~(\ref{Fig:lemc}),
the  LE as a function of $\ga$ (with the transverse field $h <1$) exhibits a sharp dip 
 near the anisotropic critical line ($\ga=0$). In contrary, when $h=1$, and $\ga$ is changed,
 we once again observe a sharp dip near $\ga=0$ which
in this case happens to be the MCP {$MC_1$} ($\ga=0,h=1$) as shown in the phase diagram Fig. (\ref{Fig:XYphase}); 
changing $\ga$ with $h=1$ implies that we are in fact probing the behavior of LE along the gapless
Ising critical line \ct{divakaran08}. It should also be emphasized here that although the spin chain
lies entirely on the critical line when the MCP is approached, we observe a substantial
dip only at the MCP which suggests that the MCP is apparently playing the role of a 
dominant critical point in determining the temporal behavior of the LE \ct{deng09}.

 \subsection{Anisotropic Critical Point (ACP)}
 
 The correlation length exponent $\nu$ and the dynamical exponent $z$ associated
 with the ACP is the same as those of the Ising transition, i.e., $\nu=z=1$ and one
 therefore expects that the behavior of LE should be similar to that close to the Ising
 transition \ct{quan06}. However, one needs to consider the fact that  at the ACP the energy gap vanishes at $\ga=0$ for
 a critical mode $k_{c} = \cos ^{-1} (-h)$.  
 
 As mentioned the decay of LE at short time is characterized by the critical exponents of the associated QCP. To calculate the
 short time behavior close to an ACP, we  define a cutoff $K_{c}$ such that only modes up to this cutoff  are incorporated
 in calculating the LE \ct{quan06} which is then given by
$
L_c(\gamma, t)= \prod _{k>0} ^{K_{c}}  L_{k}, \label{stm1}
$
 and one defines \be S(\gamma, t)=\ln L_{c} \equiv -\sum _{k>0} ^ {K_{c}} |\ln L_{k}|\ee    
 
\begin{figure}[h]
\begin{center}
\includegraphics[width=7.9cm]{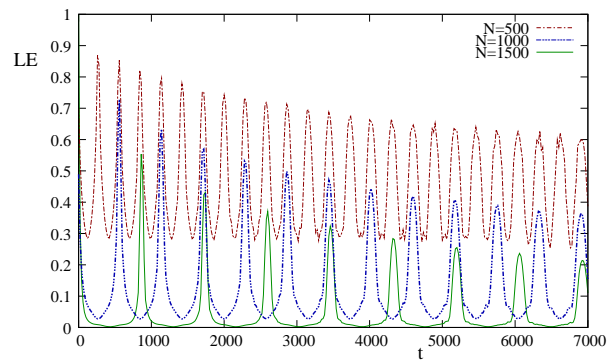}
\end{center}
\caption{ The variation of the LE as a function of time at the ACP for $\gamma = -\delta$, $h=0.5$ and $\de=0.01$. The collapse and revival of LE is indicator of a QPT and the
quasiperiod of the LE is proportional to the size $N$ of the environmental spin chain  }
\label{Fig:anisots2}
\end{figure}

\noi Expanding around the critical mode  $k_{c} $, we find
$
\sin ^2 \varepsilon _{e} ^k t \approx (\gamma +\delta)^2 k^2 t^2 $ and $\sin ^2 (2\alpha _{k}) \approx {k^2 \delta ^2}/{\{\gamma ^2 (\gamma + \delta)^2\}}$ where we have 
relabeled  $k-k_c$ as $k$;  these lead to $ S(\gamma, t) \approx - \sum _{k>0} ^{K_c} {(k \delta At) ^2}/{\gamma ^2} $. We therefore arrive at an exponential decay of
LE in the short time limit given by

\be  L_{c} (\gamma, t) \approx \exp(-\Gamma t^2) \label{sta2} \ee
where, 
$ \Gamma = {\delta ^2 E(K_{c}) k^2}/{ \gamma ^2}  $ and, 
$ E(K_{c}) = {\{4\pi ^2 N_{c}(N_{c}+1)(2N_{c}+1)\}}/{6N^2}$
(where $N_{c}$ is integer nearest to $NK_{c}/2 \pi$). From above equation (\ref{sta2}) it is clear that in this case $L_{c}$ remains invariant 
under the transformation 
$ N \rightarrow N \al$, $\delta \rightarrow \delta /\al$ and $t \rightarrow t \al$,  with $\al$ being some integer.  We now proceed to study the time evolution of LE with $h=0.5$ and $\ga=-\de$ so that the Hamiltonian $H((\ga+\de)=0)$ is critical. We observe the collapse and revival of LE  with time which is an indicator of quantum criticality as shown in  fig. ~(\ref{Fig:anisots2}). It should be emphasized  that when the  size of the spin chain ($E$)  is doubled keeping $\de$ fixed,
 the time period of collapse and
revival also gets doubled; this confirms the scaling behavior mentioned above which
is also observed at the Ising critical point \ct{quan06}.

{\bf The quasi period of oscillations can also be calculated in the following way. From Eq.~(\ref{hl1}), we find that the mode $k=k_c+2\pi/N$ gives dominant
 contribution for $t\rightarrow \infty$ so that for large $N$ limit one can expand $\varepsilon_{e}^k$ in the
form $\varepsilon_{e} ^k =h+\cos{k}
                      \approx \sqrt{1-h^2} 2\pi/N$.
We have also chosen $\gamma = -\delta$ such that  $\theta_k (\ga+\delta)=\pi/2$ which makes
\be \sin^2{2\alpha_k} = \frac{\gamma^2\sin^2{k}}{\gamma^2\sin^2{k}+(h+\cos{k})^2} \approx 1. \label{sta3}\ee
Therefore from Eqs.~(\ref{hl1}) and (\ref{sta3}), it is clear that oscillations in $L(\ga,t)$ arises due to $\sin ^2 \varepsilon _{e} ^k t$ term providing the time period
\be T=\frac{N}{2\sqrt{1-h^2}}. \label{sta4}\ee
This again shows that the time period of oscillation of the LE is proportional to the size $N$ of the environmental spin chain as shown in the Fig.~(\ref{Fig:anisots2}).
Eq.~(\ref{sta4}) also shows that the time period diverges as $h\rightarrow 1$. This originates for the fact that for $h=1$, the spin chain lies on the gapless 
Ising critical line, a situation which we are going to discuss in next sub-section.}

 We note that the decay rate $\Gamma$ scales as $\ga^{-2}$ which is consistent with the scaling given in \ct{zhang09}
since $z\nu=1$ for the transition across the anisotropic transition line also \ct{bunder99}.

\subsection{Multicritical Point (MCP)}
\begin{figure}[ht]
\begin{center}
\includegraphics[width=7.9cm]{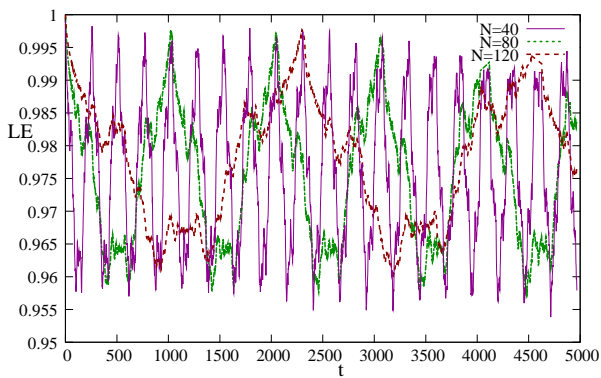}
\end{center}
\caption { The LE is shown as a function  of time  at the MCP $\gamma = -\delta$ , $h=1$ and
$\de=0.01$. The quasiperiod of the collapse and revival is proportional to $N^2$ according to the
scaling relation $ N \rightarrow N\al$, $\delta \rightarrow \delta /\al^2$ and  $t \rightarrow t \al^2$, discussed in the text. We note that the collapse and revival close to the MCP is
not a smooth function of time. }
\label{Fig:mcp1}
\end{figure}

As mentioned already, we set $h=1$, and approach the MCP by changing $\ga$ along the Ising critical line. Expanding $\sin \varepsilon _{e} ^k t$ and $\sin (2\alpha _{k})$ near MCP around $k = \pi$;
$
\sin ^2 \varepsilon _{e} ^k t \approx (\gamma +\delta)^2 k^2 t^2 $ and, $\sin ^2 (2\alpha _{k}) \approx {k^2 \delta ^2}/{\{4\gamma ^2 (\gamma + \delta)^2\}}$; $k\to (k-\pi)$, one finds
the short-time decay of LE given by

\be L_{c} (\gamma, t) \approx \exp(-\Gamma t^2) \label{stm2} \ee
where,  $ \Gamma = {\delta ^2 E(K_{c})}/{4 \gamma ^2} $ and, 
$E(K_{c}) ={ \{(1/5)N_{c} ^ 5 + (1/2)N_{c} ^4 + (1/3)N_{c} ^3 - (1/30)N_{c}\}}/{N^4} $
 (where $N_{c}$ is integer nearest to $NK_{c}/2 \pi$).  Equation (\ref{stm2}) helps in providing analytical scaling for LE i.e., $L_{c}$ is invariant 
under transformation 
$ N \rightarrow N\al$, $\delta \rightarrow \delta /\al^2$ and  $t \rightarrow t \al^2$. 
 This
scaling has to be contrasted with the scaling of $L_c$ close to the ACP presented in
the previous section. 
We note that at the MCP, the minimum energy gap scales as
$(k-\pi)^2$  so that $z=2$ whereas near an ACP , it scales linearly as $(k-k_c)$ with
$z=1$. This difference in the dynamical exponent is the reason behind different scaling
observed in the short-time limit. The collapse and revival of LE as a function of time
is shown in   Fig.~(\ref{Fig:mcp1}) for different system sizes and fixed $\de$; this confirms
the scaling observed in the short-time limit. At the
same time, we note that $\Gamma \sim \ga^{-2}$ as the exponent $z\nu=1$, even for
transition across the MCP. 

{\bf To calculate the time-period of oscillation, we again proceed using the same line of arguments given
in  section {\bf A}. At the MCP (h=1),
$ \varepsilon_{e} ^k=h+\cos{k}
                      \approx {2\pi^2}/{N^2}$
and similarly for $\ga=-\delta$, it can be shown that
$ \sin^2{2\alpha_k}  \approx 1$.
The time period of oscillations in $L(\ga,t)$ is therefore given by
$ T \approx N^2/2\pi $,
which confirms that the LE oscillates with period is proportional to $N^2$ at MCP (see Fig.~(\ref{Fig:mcp1})).
}

\section{Qubit coupled to the interaction term of the Environment Hamiltonian}

In this section we shall choose the form of the $XY$ Hamiltonian given in Eq.~(\ref{hjx1});
transforming to a new state of basis vectors defined by \ct{mukherjee07}

$$ |e_{1k}\rangle =  \sin (k/2) |0\rangle +i \cos (k/2) |k,-k \rangle$$
$$|e_{2k}\rangle = \cos (k/2) |0\rangle -i \sin (k/2) |k,-k \rangle, $$
one can rewrite the reduce $2 \times 2$ Hamiltonian (\ref{rdm})  $H_k$ in the
form
\begin{equation} 
 \left(
 \begin{array}{cc}
 J_x+J_y\cos 2k+h \cos k  & J_y \sin 2k + h \sin k\\
 J_y \sin 2k + h \sin k&  -(J_x+J_y\cos 2k+h \cos k)\\
\end{array}  \right).
\non
\label{rdm2} \end{equation}
We choose the coupling term given by
\be
H_{SE} = -  \delta \ket{e} \bra{e} \sum_{i=1}^N[ \sigma^x_i \sigma^x_{i+1}], \label{jx2}
\ee
Therefore total Hamiltonian (spin and environment) becomes
\be H_e= - \sum_{i=1}^N[(J_{x}+\delta \ket{e} \bra{e}) \sigma^x_i \sigma^x_{i+1} + J_{y} \sigma^y_i \sigma^y_{i+1} + h\sigma^z_i],  \label{jx3} \ee

The advantage of selecting such a coupling is that it enables us to explore the MCP and ACP via different paths and compare the results with the previous case, e.g.,   
if one chooses $J_x =2h_y$,  the MCP is approached along a linear path when $J_x$ is changed unlike the previous case when it is approached along the Ising critical line.

Following identical mathematical steps as described in the previous section, one can find 
that the expression of the  LE  is given by 
\be
L(J_{x}, t)=\prod _{k>0} L_{k} =\prod_{k>0}[1-\sin^2(2\alpha_{k})\sin^2(\varepsilon_ k(J_x+\de) t)] \label{jx4}
\ee
where,
$ \alpha_{k}=[\theta_k(J_x) - \theta_k(J_x+\de) ]/2 $, and
\be
\theta_{k} (J_x+\delta)= \arctan [\frac{J_{y} \sin(2k) + h \sin(k)}{J_{x}+\de + J_{y} \cos(2k) +h \cos(k)}],~~~~   \\ \label{hjx6}\
\ee
The energy spectrum given in Eq.~(\ref{hL3}) can be rewritten as
\ba
\varepsilon_k (J_x+\de) &=& [(J_{y} \sin 2k+h\sin k)^2 \non\\
&+ &(J_{x}+\de+J_{y}\cos 2k+h\cos k)^2]^{1/2} .\label{hjx8}
\ea

\begin{figure}
\begin{center}
\includegraphics[height=6.6cm,width=8.8cm]{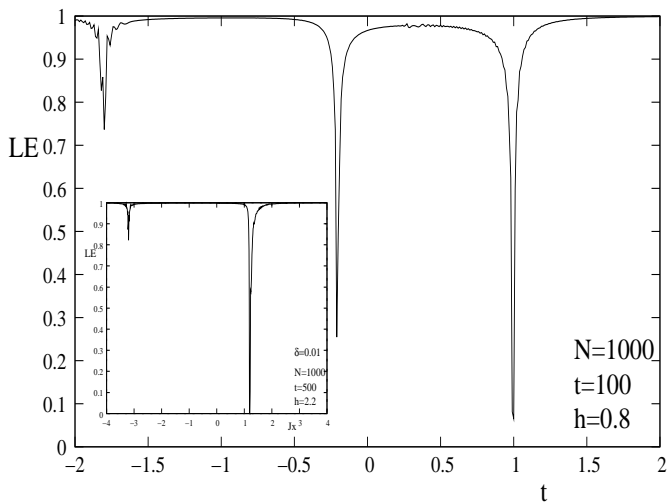}
\end{center}
\caption{ The  LE plotted as a function of  $J_{x}$ shows dips around the  Ising critical points  ($J_{x}= -0.2, -1.8$) as well as the ACP ($J_{x}=1$) with  $J_{y}=1$ and $h=0.8 (<2J_y)$. Inset shows that for $h= 2.2(>2J_y) $, there are dips only at the Ising critical points as
the variation of $J_x$ does not take the spin chain across the anisotropic critical line.
 }
\label{Fig:hless}
\end{figure}

\begin{figure}
\begin{center}
\includegraphics[width=6.4cm]{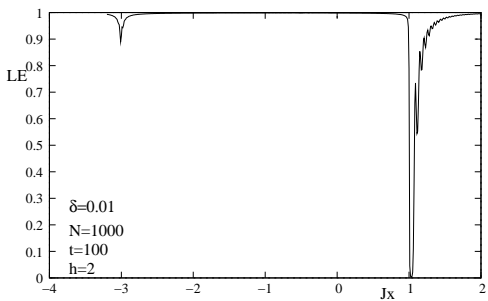}
\end{center}
\caption{The interaction $J_x$ is varied with $h$ fixed to $h=2J_y=2$. The LE  shows a dip at Ising critical point($J_{x}= -3$) and also at the  MCP ($J_{x}=1$). }
\label{Fig:heq}
\end{figure} 
Let us first explore the   LE close to different critical points; refereeing to fig~(\ref{Fig:hless})  and (\ref{Fig:heq}), we find that there is a sharp dip in LE wherever the
parameters values are such that the system is close to a critical point. For example,
in Fig.~(\ref{Fig:hless}), we have varied $J_x$ keeping $h$ and $J_y$ fixed and $h<2J_y(=2)$ such that we observe dips at two Ising critical points and also at the anisotropic critical point; for $h>2J_y$, in contrast, one observes dips only at the Ising
critical points as the anisotropic transition point is not crossed in the process of changing
$J_x$. For $h=2J_y$, one observes dips at the Ising critical point and the MCP as
$J_x$ is varied (Fig.~(\ref{Fig:heq})). Equipped with these observations, we now proceed
to study the short time decay of LE close to these critical points. 

\subsection{Short time behavior} 

\begin{figure}[ht]
\begin{center}
\includegraphics[height=6.0cm,width=8.6cm]{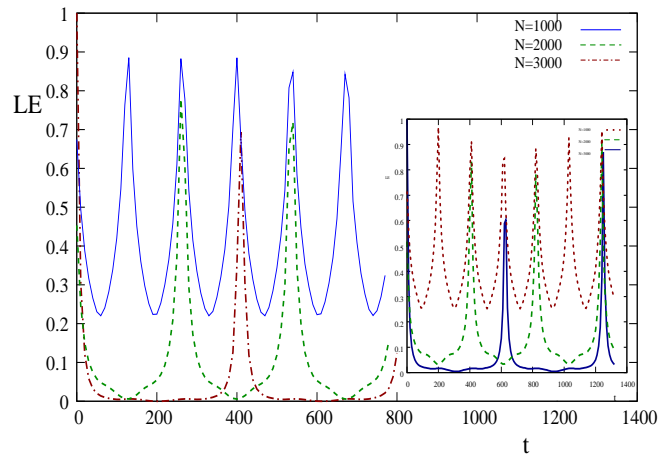}
\end{center}
\caption{ The collapse and revival of the LE  at the  ACP  ($ J_{x} = 1-\delta$ , $h=0.8, J_y=1$). The inset shows the same behaviour at the Ising critical point ($h=2.2, J_x=1-\de$ and
$J_y=1$). In both the cases $\de=0.01$ }
\label{Fig:anisola}
\end{figure}

\begin{figure}[ht]
\begin{center}
\includegraphics[width=7.9cm]{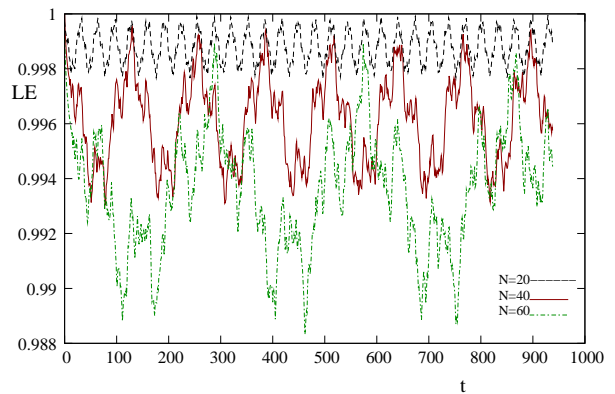}
\end{center}
\caption{The time variation of LE  at MCP ($J_x=1-\de$, $h=2=2J_y$) is shown . The quasiperiod of LE is again proportional to  $N^2$ as reported in Sec. II.  }
\label{Fig:anisola1}
\end{figure}

Near the critical points and the MCP, we have identical short-time  behavior  and scaling 
with respect to $N$, $\de$ and $t$ as already reported in the previous section. These are corroborated by numerical estimation
of collapse and revival close to the critical and the multicritical point as shown in
Figs.~(\ref{Fig:anisola}) and   (\ref{Fig:anisola1}). This confirms that the scaling of 
LE does not depend on how the central spin is coupled to  the environment  rather it is 
hypersensitive to the  proximity to a critical point of the environment.
  
\subsection{Close to the MCP}

\begin{figure}[ht]
\begin{center}
\includegraphics[width=7.9cm]{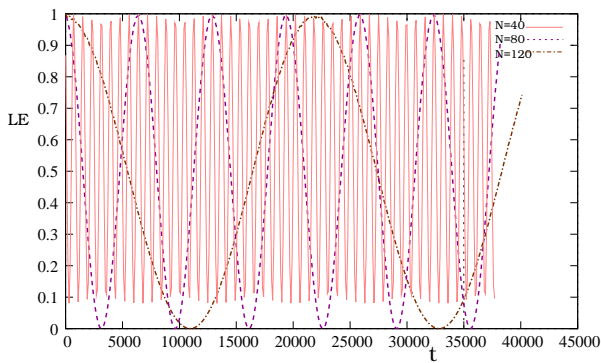}
\end{center}
\caption{  Collapse and revival of the LE at a quasicritical point  ($ J_{x} = 1-\delta+4\pi ^2/ N^2 $ , $ h=2 $ and $ J_{y}=1 $) as defined in the text. The quasiperiod scales with
$N^3$ in contrast to $\sim N^2$ at the MCP. This is consistent with the scaling
$N \rightarrow N\al$, $\delta \rightarrow \delta /\al^3 $ and  $t \rightarrow t\al^3$. }
\label{Fig:jxMCPt}
\end{figure}

It is well known that for a finite $XY$ spin chain, there exist quasicritical points on the 
ferromagnetic side  close to the MCP; the energy gap is locally minimum at these quasicritical points and it scales $k^3$ 
in contrast to the scaling $k^2$ at the MCP. In the limit of $N \to \infty$, all these
quasicritical points approach the MCP. These
quasicritical points and exponents associated with them have been found to dictate the
scaling of the defect density following a slow quench across the MCP
\ct{deng09,mukherjee10}  and also the scaling
of fidelity susceptibility close to it  \ct{mukherjee11}. We shall
now explore the collapse and revival of the LE fixing the parameters such that 
$J_x + \de$ is right at a quasicritical point. For modes  $k \approx \pi$, one can
use the simplification,
 $
\sin ^2 \varepsilon _{e} ^k t \approx (J_{x} + \delta - J_{y})^2 t^2 $ and, $\sin ^2 (2\alpha _{k}) \approx {4J_{y} ^2 k^6 \delta ^2}/{\{(J_{x}-J_{y})^2 (J_{x} + \delta -J_{y})^2\}}$. 
We therefore get a similar exponential decay of the LE
$ L_{c} (J_{x}, t) \approx \exp(-\Gamma t^2) $ with

\be
\Gamma = \frac{4 J_{y} ^2 \delta ^2 E(K_{c})}{(J_{x}-J{y})^2} \label{jx10} ~~{\text and} ~~
 E(K_{c}) =\frac { A(N_c)}{N^6}, 
               \label{jx11} \ee
where 
$A(N_c) =(1/7)N_{c} ^7 + (1/2)N_{c} ^6 + (1/2)N_{c} ^5 - (1/6)N_{c} ^3 + (1/42)N_{c}$, and as defined previously, 
$N_{c}$ is integer nearest to $NK_{c}/2 \pi$. The above equation (\ref{jx11}), shows a very interesting scaling behavior of the LE 
$N \rightarrow N\al$, $\delta \rightarrow \delta /\al^3 $ and  $t \rightarrow t\al^3$ which
is different from the scaling observed at the MCP.
{\bf Simillar to previous cases, at the quasicritical point
$ \varepsilon_{e} ^k=h+\cos{k}
                      \approx {16\pi^3}/{3N^3}$ with  $h=2J_y=2$; 
for $J_x=1-\delta+4\pi^2/N^2$ and large N, it can be easily shown that
$ \sin^2{2\alpha_k}  \approx 1$.
The time period of oscillations in $L(J_x,t)$ is therefore given by
$ T \approx {N^3}/{16\pi^2} $, which 
verifies the fact that LE oscillates with period proportional to $N^3$ at quasi critical point (as shown in Fig.~(\ref{Fig:jxMCPt})).
}

 The collapse and revival of the LE
as a function of time supports the scaling behavior analytically obtained in the short
time limit (see Fig.~(\ref{Fig:jxMCPt})). Comparing Eq.~(\ref{jx11}) with the form of decay rate $\Gamma$ given in Eq.~(\ref{sta2}), we find that
in both the cases $\Gamma \sim 1/\ga^2$; this is because at a quasicritical
point  one can define an effective dynamical exponent $z_{qc}=3$ $\nu_{qc}=1/3$ such
that $\nu_{qc}z_{qc}=1$ \ct{deng09,mukherjee10}. Moreover,  we
find that the quasiperiod scales as $N^{z_{qc}}$.

\section{Conclusion}

In this paper, a spin-1/2 (qubit) is coupled to the environment which is chosen to
be a spin-1/2 $XY$ spin chain and  the temporal behavior  of the LE is studied. The coupling
is done in such a way that  enables us to study the LE close to the ACP
as well as the MCP of the phase diagram and these points are approached in different
fashions, e.g., the MCP is approached along the Ising critical line in Sec. II while in Sec. III
it is approached following a linear path. We find that close to the ACP, the evolution
of the LE is identical to that reported in the ref. \ct{quan06}. However, around the MCP,
we observe that the  quasiperiod of the collapse and revival of the LE
as a function of time scales as $N^2$ where $N$ is the size of the environmental
spin chain. We attribute this to the fact that the dynamical exponent $z$ associated
with the MCP is two. To justify this conjecture, we have estimated  the scaling of the decay
rate $\Gamma$ and also the period of the collapse and revival of the LE at a quasi-critical
point on the ferromagnetic side of the MCP. We find that quasiperiod scales as $N^3$.
It should be noted here that at the quasicritical point, the minimum gap scales with
the system size as $N^3$ and hence one can define an equivalent dynamical
exponent $z_{qc}=3$ \ct{mukherjee10}. In Sec. II, even though the MCP is approached along
a gapless critical line, a sharp dip in the LE is observed only around the MCP where the decay of
 energy gap with the system size is faster ($\sim 1/N^2$) with respect to that near the
Ising or anisotropic critical point. We observe that the collapse and revival of the LE at MCP is not a smooth function of time which is attributed to the fact that in Sec II the spin chain is
always close to  the Ising critical line whereas  in Sec. III quasicritical points  are likely
to influence the temporal evolution of the LE. These quasi-critical points, on the other hand, are expected to be related to the proximity to the critical line of the finite-momentum anisotropic transition.

Although we have studied an integrable spin chain reducible to direct product of two-level system, our
studies indicate the possibility of some interesting scaling behavior. We see that in
all the cases studied here, the LE decays exponentially close to the critical point in the
short time limit
with the decay rate $\Gamma$ scaling as  $\Gamma \sim \lambda^{-2z\nu}$ i.e., our studies
support the scaling proposed in \ct{zhang09} based on  perturbative calculations and
a Landau-Zener argument. Moreover, we find the quasiperiod of the collapse and revival
of the LE at the critical point scales as $N^z$; we note that the dynamical  exponent $z$
 determines how does  the minimum energy gap vanishes with increasing system size
 ($\sim N^{-z}$) at the QCP.  At a quasicritical point
the  effective dynamical exponent $z_{qc}=3$ is found to determine 
the scaling of the quasiperiod of collapse and revival with the system size.

Finally, we comment on the decoherence of the central spin during time evolution which is calculated using its
reduced density matrix \ct{quan06}. The off-diagonal terms of the reduced density matrix is given by
$c_g^{*}c_e d(t)$ and its hermitian conjugate where the decoherence factor $d(t)$ is connected to the LE through the
relation $L(t)=|d(t)|^2$ \ct{damski11}. The vanishing of the LE around to the QCP therefore implies
a complete loss of coherence and therefore the qubit makes transition to a mixed state even though initial state
is chosen to be pure. On the other hand, away from the QCP LE stays close to unity, thus the purity of the qubit
state is retained. Our studies reveal that close to the MCP, $\Gamma \sim 1/N^2$ in the short time limit, implying a faster loss of
coherence with the increasing system size when the environment $E$ is close to a MCP than when it is close to a QCP. The loss
is even faster when the spin chain sits at a quasicritical point close to the MCP. This faster
loss of coherence with the system size, we believe, is a note-worthy observation.

\begin{center}
{\bf Acknowledgements}
\end{center}

 AD acknowledges CSIR, New Delhi, India, for financial support through research project and SS acknowledges CSIR, New Delhi,
for junior research fellowship.

\end{document}